\documentclass[twocolumn,amsmath,amssymb,showpacs]{revtex4}
\usepackage{subfigure}
\usepackage{wrapfig}
\usepackage{graphicx}
\usepackage{dcolumn}
\usepackage{bm}
\usepackage{amsmath}
\usepackage{amsfonts}
\usepackage{amssymb}
\usepackage{color}
\usepackage{mathrsfs}
\makeatletter

\begin{document}
\title{Slip Morphology of Elastic Strips on Frictional Rigid Substrates}
\author{Tomohiko G. Sano$^{1,2}$, Tetsuo Yamaguchi,$^3$ and Hirofumi Wada$^1$}
\affiliation{
$^1$Department of Physical Sciences, Ritsumeikan University, Kusatsu, Shiga 525-8577, Japan\\
$^2$Research Organization of Science and Technology, Ritsumeikan University\\
$^3$Department of Mechanical Engineering, Kyushu University, 744 Motooka, Nishi-ku, Fukuoka 819-0395, Japan
}
\begin{abstract}
The morphology of an elastic strip subject to vertical compressive stress on a frictional rigid substrate is investigated by a combination of theory and experiment. We find a rich variety of morphologies, which---when the bending elasticity dominates over the effect of gravity---are classified into three distinct types of states: pinned, partially slipped, and completely slipped, depending on the magnitude of the vertical strain and coefficient of static friction. We develop a theory of elastica under mixed clamped--hinged boundary conditions combined with the Coulomb--Amontons friction law, and find excellent quantitative agreement with simulations and controlled physical experiments. We also discuss the effect of gravity in order to bridge the difference in qualitative behaviors of stiff strips and flexible strings, or ropes.
Our study thus complements recent work on elastic rope coiling, and takes a significant step towards establishing a unified understanding of how a thin elastic object interacts vertically with a solid surface.
\end{abstract}
\maketitle

{\itshape -Introduction:} Contact between slender objects gives rise to complex structures and behaviors in nature~\cite{darcy_thompson,ghosal_prl_2012,dumais_forterre_2012,wolpert, hohn_prl_2015,sachs_book,goriely_prl_1998,engelberth_2003,jin_jeb_2013,alexander_katz_2014,menzel_2011,dietler_2001,massa_gilroy_2003,moushausen_2009,mahadevan_friction_2015}, including DNA ejection from bacteriophages~\cite{ghosal_prl_2012}, the folding of sheet-like tissues in developmental biology~\cite{wolpert, hohn_prl_2015}, and the coiling of plant tendrils or roots~\cite{sachs_book,goriely_prl_1998,engelberth_2003}. 
Examples in daily life~\cite{goldstein_prl_2012,callan_jones_prl_2012,jawed_prl_2015,habibi_prl_2007,p_reise_pnas_2014,ribe_pre_2003,brun_prl_2015,mahadevan_proc_1996,rug_1,rug_2} include hair brushing, arranging pony tails~\cite{goldstein_prl_2012}, applying gift-wrap ribbons~\cite{callan_jones_prl_2012}, tying shoelaces~\cite{jawed_prl_2015}, rucks in a rug~\cite{rug_1,rug_2}, coiling elastic or liquid ropes~\cite{ribe_pre_2003,habibi_prl_2007,p_reise_pnas_2014,brun_prl_2015,mahadevan_proc_1996}, or the use of polymer brushes~\cite{klein_nature_1994}, biomimetics~\cite{gecko_1,gecko_2,gecko_3,gecko_4,gecko_5} and coiled tubing in industry~\cite{miller_jss_2015}. 
Since frictional effects~\cite{persson_book,otsuki_matsukawa,alarcon_2016,stoop_2008,rubinstein_2004} play an important role when slender objects are in contact with each other~\cite{mahadevan_friction_2015}, the interplay between friction and the elasticity of thin objects is currently a central topic in this field of research.

\begin{figure}[h]
\begin{center}
\includegraphics[scale = 0.43]{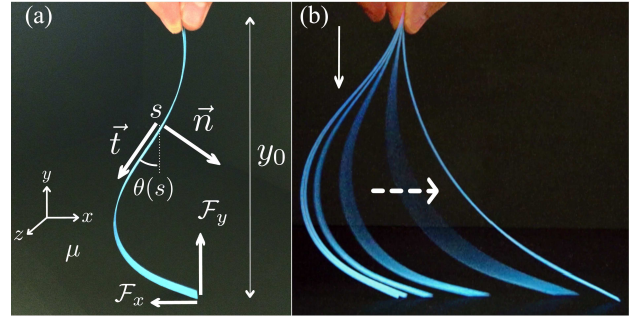}
\caption{(Color online).~Typical morphology of a strip on a solid surface. 
(a) Geometry of our system and definition of the key variables. 
$\vec{t}$ and $\vec{n}$ are the unit tangent and normal vectors of the strip centerline, respectively. $\theta(s)$ represents the angle of $\vec{t}$ measured from the $y$ axis.
The strip either slips or is pinned, depending on the force from the substrate, $\vec{\mathcal{F}}=({\mathcal F}_x,{}{\mathcal F}_y)$, coefficient of static friction, $\mu$, and vertical height $y_0$. (b) Photograph of the slip motion of a strip (for illustrative purposes only).}
\label{setup}
\end{center}
\end{figure}

A fundamental process common to a variety of the problems listed above is the postbuckling behavior of an elastic
strip~\cite{audoly_book,landau_elasticity,euler_buckling1,wang_review_1986,he_etal_1997,plaut_2011}, that is subject to a vertical compressive stress on a rigid substrate [Fig.~\ref{setup}(a)]. 
Initially, the strip takes the form of a planar elastica, but upon further compression, its free tip may slip [Fig.~\ref{setup}(b)]. 
The direction of this slippage is opposite to the direction of the initial buckling (which is determined by spontaneous symmetry breaking), as the slip acts to reduce the overall curvature of the strip. 
Despite the familiarity, simplicity, and fundamental importance of this prototypical phenomenon, its underlying physics remains unclear thus far. 
For example, several unanswered questions are:
when and how does the strip slip, what factors determine the slip length, and what are the possible resultant forms of the elastica?
To answer these basic questions, it is necessary to disentangle the complex interplay between elasticity, geometry, friction, and gravity.

\begin{figure}[t]
\begin{center}
\includegraphics[scale = 0.93]{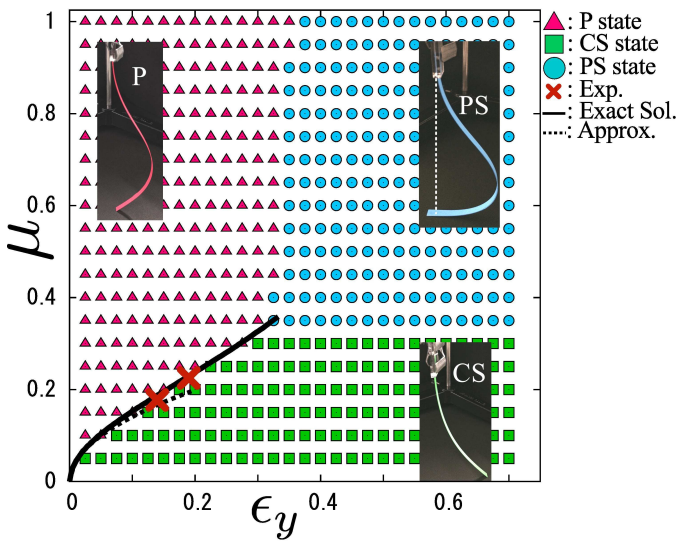}
\caption{Phase diagram of the equilibrium planar shapes of a strip in the $(\epsilon_y,\mu)$ parameter space, constructed from numerical simulations for compressive protocols in the absence of gravity. 
Two different experimental data points are plotted as the thick red $\times$ symbols.
The solid line represents the theoretical prediction based on the exact solution for the elastica curve, and 
the dashed line is its approximation given by Eq.~(\ref{critical_mu}).
Note that the theoretical curve ends when the areal contact begins and the elastica solution no longer exists.
}
\label{diagram}
\end{center}
\end{figure}

In this letter, we investigate the above-outlined problem using numerical, analytical, and experimental approaches.
The frictional interaction between the strip's tip and the surface of the substrate is modeled according to the Coulomb--Amontons law~\cite{persson_book,otsuki_matsukawa,alarcon_2016,stoop_2008,rubinstein_2004}, which states that the tip of the strip remains stationary if the frictional force from the substrate, $\vec{\mathcal F}=({\mathcal F}_x, {\mathcal F}_y)$, satisfies
\begin{eqnarray}
|{\mathcal F}_{x}|\leq\mu\mathcal{F}_y,
\label{eq:CA-law}
\end{eqnarray}
where $\mu$ represents the coefficient of static friction. 
Equation~(\ref{eq:CA-law}) suggests that the instantaneous shape of an elastica determines its own boundary condition.
This feature is a particular characteristic of our system and differs from the behavior in a standard setup employed in previous elastica problems~\cite{wang_review_1986,he_etal_1997,plaut_2011}. 
First, we numerically investigate the planar deformations of a strip in the absence of gravity by changing the values of the coefficient of static friction, $\mu$, and the height of the strip, $y_0$, and classify the deformations into three distinct states explained below. 
These morphologies, as well as the shape transitions between them, are confirmed by our experiments. 
An analytic model based on geometrically exact Kirchhoff rod equations combined with the static friction law is then developed, which accurately predicts the onset of slip events observed in the simulations. 
Finally, we explore the effect of gravity to discuss how our system may approach those studied in the context of elastic rope coiling~\cite{habibi_prl_2007,p_reise_pnas_2014}.

{\itshape -Simulations:} 
To investigate the planar deformation of a strip with slip in a geometrically nonlinear regime, we performed systematic numerical simulations using a discrete analog of the continuum elastica model~\cite{discrete_model}. 
The centerline of the strip was discretized into a chain of $N$ spheres with a bond length, $b_0$.
As we are interested only in the equilibrium shape of the chain, we model the evolution of the positions of each sphere according to the overdamped equations of motion, where both stretching and bending elastic forces act on the chain of spheres. 
Further details are given in the Supplementary Materials (SM)~\cite{supp,comment2}.

The top end of the strip is clamped along the vertical ($y$) axis.
Generally, when the strip's tip is in contact with the substrate, the tip experiences forces and moments from the substrate but is otherwise free. 
In this study, we assume moment-free boundary conditions at the tip, even when in contact with the substrate. 
The force from the substrate is determined according to Eq.~(\ref{eq:CA-law})~\cite{persson_book}.
Once $|\mathcal{F}_{x}|$ exceeds $\mu\mathcal{F}_y$, the kinetic friction force, $\mu_{\rm k} \mathcal{F}_y$, takes over, acting to oppose the continued slipping of the strip.
As soon as the tangential force falls below this threshold, the static friction sets in again.
We confirmed that our results are insensitive to a precise static--kinetic switching protocol (See our SM for further details~\cite{supp}).

Our strip is initially a vertically aligned straight line with sufficiently small random displacements along the $x$ direction only, which induce the initial buckling.
We change the position of the clamped end at a given speed, $u$, so that the strip of initial length, $L$, is pushed against the substrate from directly above, until its height reaches a given value, $y_0{}(< L)$. 
The stretching modulus is set to a sufficiently large value, in order to restrict the typical variation in arc length to within a few percent.
This allows us to compare our simulation results with analytical predictions for an inextensible elastica, which are described below.
Similarly, the velocity of the clamped end is chosen to be sufficiently small when compared to the bending relaxation time~\cite{supp}, in order to minimize any protocol-dependent kinetic effects. 
Throughout this work, we use a chain of $N=30$ spheres, and a kinetic and static frictional coefficient ratio of, $\mu_{\rm k}/\mu=0.8$, which is valid for typical surfaces.

{\itshape -Slip morphology of elasticas:}
We consider a planar bending deformation of a straight strip of length, $L$, characterized by a radius of curvature, $R$.
The bending torque is $EI/R$, where $E$ is Young's modulus, and $I$ is the moment of inertia of the strip.
Since the typical displacement perpendicular to the strip axis is $L^2/R$, the gravitational torque acting on the strip is ${\rho}gL^3/R$, where $\rho$ represents the mass {\it per unit length} along the strip-centerline.
Balancing the two torques provides a so-called ``gravito-bending length''~\cite{p_reise_pnas_2014,wang_review_1986},
\begin{eqnarray}
L_g&=&\left(\frac{EI}{{\rho}g}\right)^{1/3}.
\label{eq:L_g}
\end{eqnarray}
The dimensionless parameter, $L/L_g$, quantifies the relative importance of gravity to the elasticity. 
For $L/L_g\gg1$, the strip is significantly deformed by gravity (i.e., by its own weight), and this scenario has been extensively studied previously~\cite{p_reise_pnas_2014,wang_review_1986}.
Here, we are interested in the opposite limit, $L/L_g\ll1$, in which the behavior of a stiff strip is effectively studied by neglecting gravitational body forces.
Our systematic numerical investigations in this regime are summarized in a phase diagram in Fig.~\ref{diagram}.
The shapes are classified as pinned (P) (for large $\mu$ and small $\epsilon_y{}\equiv 1 -y_0/L$), partially slipped (PS) (for large $\mu$ and large $\epsilon_y$), and completely slipped (CS) (for small $\mu$ and large $\epsilon_y$) states. 
If the tip remains stationary, it is said to be a P state; if it slips, and the final shape has an inflection point, it is a PS state; otherwise, it is a CS state.
The phase diagram in Fig.~\ref{diagram} is constructed according to this protocol~\cite{supp}.
We find clear boundaries between the three regions in Fig.~\ref{diagram}, which we now rationalize using the exact theory of elastica and scaling arguments.

{\itshape -Phase boundaries:} 
The diagram in Fig.~\ref{diagram} suggests that between the P and CS states, the coefficient of static friction assumes a critical value, which depends on the vertical strain $\epsilon_y$, i.e., $\mu_c=\mu_c(\epsilon_y)$. 
The geometry of our analytic theory is shown in Fig.~\ref{setup}(a), where the unit tangent is parameterized as $\vec{t}(s)=(\sin\theta(s),-\cos\theta(s))$, using the variable $\theta(s)$, where $s$ is the arc length measured from the clamped top, $s=0$.
The relevant boundary conditions are thus written as, $\theta(0)=0$, and, $\theta'(L)=0$, where the prime symbol represents the derivative with respect to $s$. 
This latter boundary condition suggests that no external moment is applied at the end of the strip.
We now let $\vec{F}(s)$ and $\vec{M}(s)$ be the internal force and moment, respectively, over the cross section of a strip at the position $s$, and which are exerted by
the section of the strip with an arc length greater than $s$, on the section of the strip with an arc length less than $s$~\cite{powers_rmp_2011,nizette_goriely_1999}.
In the absence of any external forces and moments, the force balance of an elastica is described by the Kirchhoff rod equations~\cite{audoly_book,landau_elasticity}, $\vec{F}'(s)=0$, and $\vec{M}'(s)+\vec{t}(s)\times\vec{F}(s)=0$, and the linear constitutive relation $\vec{M}(s)=EI\theta'(s)\hat{e}_z$. Both tangential and vertical external forces must be applied at the clamped end, i.e., $\vec{F}(0)=(\mathcal{F}_x, \mathcal{F}_y)$, where $\mathcal{F}_x, \mathcal{F}_y$, are yet to be
determined. Solving the force-balance equation with this condition and substituting it into the momentum balance equation leads us to
the shape equation for $\theta(s)$~\cite{he_etal_1997,plaut_2011}, which may be written as 
$EI\theta''(s)=-\mathcal{F}_x\cos\theta(s)-\mathcal{F}_y\sin\theta(s)$.
Although the equation for $\theta(s)$ can be analytically solved using elliptic integrals~\cite{mitaka_acta_mechanica_2007,abramowitz,inclined_paper}, the result for small strain, $\epsilon_y\ll1$, is useful for our aim here. 
Since $|\theta|\ll1$ for $\epsilon_y\ll1$, the approximate solution becomes $\theta(s) = ({f_x}/{f_y})\left\{\sqrt{f_y}\sin(\sqrt{f_y}s/L) + \cos(\sqrt{f_y}s/L)-1\right\}$, with $f_{x}\equiv{\mathcal F}_{x}L^2/EI$ and $f_{y}\equiv{\mathcal F}_{y}L^2/EI$.
Furthermore, this expression must satisfy the shape constraints for a pinned strip, given by
$x(L)=\int_{0}^{L}ds\sin\theta(s)=0\label{const_x0}$ and 
$y(L)=-\int_{0}^{L}ds\cos\theta(s)+y_0=0\label{const_y0}$, 
from which we obtain the equations given by,
$\tan\sqrt{f_y}=\sqrt{f_y}$, and $f_x/f_y=\sqrt{4\epsilon_y/f_y}$~\cite{supp}.
Solving these numerically, we find the value, $\sqrt{f_y}\simeq4.4934$. 
The shape reconstructed from the approximate solution is shown in Fig.~\ref{sche_rod_bend_his}(a), and describes the configurations observed in our simulations and experiments (outlined below) quite well.
Combining the above result with the slip condition for $\vec{\mathcal{F}}$ in Eq.~(\ref{eq:CA-law}), we arrive at
\begin{equation}
\mu_c(\epsilon_y) = \frac{\mathcal{F}_x}{\mathcal{F}_y}\simeq0.4451\sqrt{\epsilon_y}\label{critical_mu}.
\end{equation}
Again, this approximate solution matches the exact elliptic function solution ~\cite{inclined_paper} quite well for $\epsilon_y\ll 1$ and is in excellent agreement with the numerical data, as seen in Fig.~\ref{diagram}.

The boundary between the P and PS states in the phase diagram [Fig.~\ref{diagram}] suggests that a maximum vertical strain $\epsilon_{y}^{\rm max}$ exists for the P state, which is {\it independent} of $\mu$, for large $\mu$.
This observation is corroborated on the basis of the following simple argument. 
Once $\theta(L)$ reaches $\pi/2$, a P configuration may be permanently stabilized because 
the contact area between the strip and the surface increases with any further compressive force.
Assuming then that the shape of the bent strip is close to that of a semicircle of radius $R_{\rm eff}$ [see Fig. \ref{sche_rod_bend_his}(b)], and regarding $\pi R_{\rm eff}\simeq L$, we obtain $\epsilon_{y}^{\rm max}\simeq(L-2R_{\rm eff})/L\simeq1-2/\pi\simeq0.363$. This rough argument yields a surprisingly good prediction for the position of the phase boundary between the P and PS states in Fig.~\ref{diagram}.

\begin{figure}[b]
\begin{center}
\includegraphics[scale = 0.68]{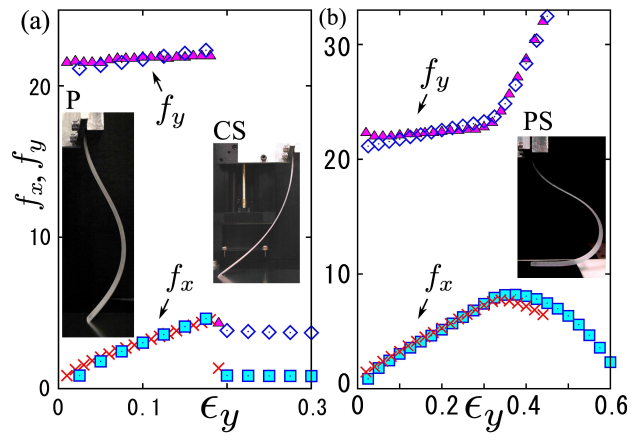}
\caption{(Color online). Rescaled total tangential ($f_x$) and normal ($f_y$) forces acting on a strip by the substrate, measured in simulations and experiments and plotted as a function of the vertical strain $\epsilon_y$. 
(a) $f_x$ ($\times$) and $f_y$ ($\triangle$) from experiment with an aluminum surface.
$f_x$ ($\square$) and $f_y$ ($\diamond$) from simulations with $\mu=0.225$.
(b) $f_x$ ($\times$) and $f_y$ ($\triangle$) from experiment with a rubber surface.
$f_x$ ($\square$) and $f_y$ ($\diamond$) from simulations with $\mu=0.40$.
Insets in (a) and (b) are the experimental snaphots.}
\label{fig:force_strain_relation}
\end{center}
\end{figure}

{\itshape -Experiments:}
To verify our main theoretical findings, we conducted controlled physical experiments using a slender elastic strip made of polyvinyl chloride (PVC) of length, $L = 150$ mm, width $10$ mm, and thickness $1$ mm.
The Young's modulus of such a PVC strip is known to be $E=2.4$--$4.1\times10^9$ Pa.
The bottom and side faces of our strip were polished with sandpaper to add some surface roughness, and two types of substrates---an aluminum plate and a carbon-filled natural rubber sheet---were used to vary the frictional coefficients 
in a controlled manner. 
In the experiments, the head of the $z$--stage clamping the PVC strip moved downward sufficiently slowly by a distance of 1--2\% of the strip's length.
At every step, the clamping end was kept fixed for 30 seconds so that the strip attained its equilibrium position, after which the {\it total} tangential and horizontal forces that the strip exerted on the substrate were measured.
See our SM for full experimental details~\cite{supp}.

In Figs.~\ref{fig:force_strain_relation} (a) and (b), the experimental force vs. strain relations are plotted, together with those predicted from our simulations.
The forces are rescaled in units of $EI/L^2$. 
We find an excellent agreement between simulation and experiment, from which we could estimate $\mu\simeq0.225$, for the case in Fig.~\ref{fig:force_strain_relation}(a)~\cite{comment}. This particular experimental point, $(\epsilon_y, \mu) = (0.19, 0.225)$, as well as data from another experiment, $(\epsilon_y, \mu) = (0.14, 0.175)$, are superposed on the diagram in Fig.~\ref{diagram}. The two data points sit exactly on the phase boundary between P and CS predicted from simulations and our theoretical analysis.

In the experiment with the aluminum plate [Fig.~\ref{fig:force_strain_relation} (a)], we observed the transition to the CS state.
In this case, the tangential and normal forces increase as the strip buckles, and at the slip transition, these forces experience an abrupt and discontinuous decrease in strength.
In contrast, the PS state occurs in the experiments with the rubber substrate [Fig.~\ref{fig:force_strain_relation} (b)], where the force curves are distinct from those in the CS case.
Across the transition to the PS state, the normal force starts to increase in magnitude, while the tangential force begins to decrease continuously.
A closer look at this event reveals that the partial slip involves the onset of the areal contact between the strip and the substrate.
This geometric transition is continuous, and acts to reduce the tangential tension, while increasing the normal force significantly.

\begin{figure}[t]
\begin{center}
\includegraphics[scale = 0.95]{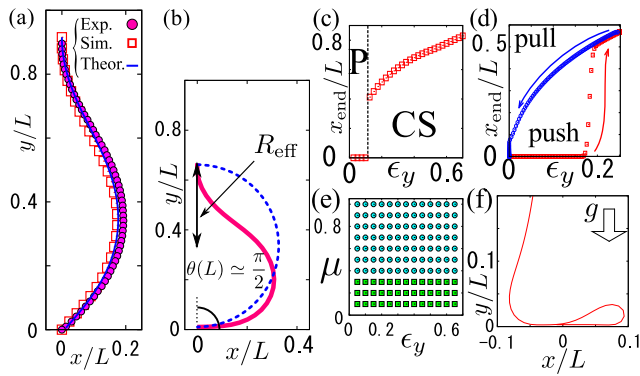}
\caption{(Color online). (a) Comparison of simulation ($\square$) for $\mu = 1.0$, experiment ($\circ$), and analytic theory (solid line) for P states with $\epsilon_y = 0.10$. (b) Critical P configuration (from the simulation) close to the P--PS boundary, $(\epsilon_y, \mu) = (0.35, 1.0)$. (c) Rescaled slip distance $x_{\rm end}/L$, plotted as a function of $\epsilon_y$, obtained from the simulations for $\mu = 0.2$ during the compression process. (d) $x_{\rm end}/L$, plotted as a function of $\epsilon_y$, with $\mu = 0.2$ for compression (squares) and reverse (circles) processes. (e) Phase diagram for the reverse protocols. (f) Self-folding shape for $L/L_g\simeq6$, and $N=100$.}
\label{sche_rod_bend_his}
\end{center}
\end{figure}

{\itshape -Hysteresis:}
In Fig.~\ref{sche_rod_bend_his}(c), the position of the free end, or the slip distance, $x(s=L)=x_{\rm end}$, obtained from our simulations is plotted as a function of $\epsilon_y$ for $\mu = 0.2$~\cite{supp}.
A discontinuous change in $x_{\rm end}/L$, at the transition from the P to the CS state appears in Fig.~\ref{sche_rod_bend_his}(c), whereas $x_{\rm end}$ changes continuously from zero at the transition from the P to the PS state (data not shown).
Furthermore, Fig.~\ref{sche_rod_bend_his}(d) shows that the trajectory of $x_{\rm end}$ in the reverse process differs considerably from that of the compression process, revealing a distinct hysteresis in the cyclic process.
In particular, the strip never returns to the P configuration in the reversed process. We show the phase diagrams generated by the reverse processes in Fig.~\ref{sche_rod_bend_his}(e), where no P state exists.
Such protocol-dependent hysteretic behavior, or multi-stability, is a direct consequence of the friction law, and has also been found in granular experiments under shear~\cite{nasuno_prl_1997}.

{\itshape -Effect of gravity:} We now discuss the effect of gravity by changing the dimensionless ratio, $L/L_g$, while fixing $(\epsilon_y, \mu)=(0.15, 0.2)$, so that the strip is in the P configuration for $g=0$. 
As $L/L_g$ is increased, the strip tends to sag, while for $L/L_g = O(1)$, i.e., when the bending is comparable to the gravity, the free end slips easily because the effect of gravity acts to increase $\theta(L)$, and the horizontal force $\mathcal{F}_x$ also increases. 
However, the resulting shape is distinct from that of the CS state because the strip interacts with the substrate by areal contact rather than point contact. 
As $L/L_g$ becomes larger, [$L/L_g\simeq6$ in Fig.~\ref{sche_rod_bend_his}(f)], the strip folds and loops back on itself. 
This folding is analogous to the planar version of an elastic rope coiling.
Actually, using the same parameter set, we can reproduce a realistic coiling shape in our three-dimensional simulation of a twist-free elastic string~\cite{supp}. 
Therefore, our investigation here manifests the following physical scenario about the initiation of the coiling.
First, the free end slips partially immediately upon contact with the substrate. Subsequently, the contact length increases monotonically as the string sags, which significantly reduces the tangential tension, and prevents the string from slipping further. 
This effectively confines the sagging string to a localized position, eventually leading to the characteristic coiling.

{\itshape -Conclusion:} 
We investigated the planar slip configurations of an elastic strip pushed onto a frictional rigid substrate.
Combining numerical, analytical, and experimental approaches, we revealed the fundamental aspects of this problem in the limit of weak gravitational effects, and quantified the relative importance of the system's geometry, elasticity, friction, and gravity.
The framework presented here could be applied to a number of biophysical phenomena across different scales, including membrane-bound actin polymerization in cell motility~\cite{Mogilner_1996,Daniels_2006}, gravity-guided intrusion of plant roots in soil~\cite{massa_gilroy_2003,moushausen_2009}, and contact mechanics of the adhesive hairs in geckos' toe pads~\cite{gecko_1,gecko_2,gecko_3,gecko_4,gecko_5}. In all of these examples, bending of thin elastic objects against rigid or flexible substrates occurs, determining the overall behavior. This suggests a profound connection between the mechanical processes and the specific biological functions in those systems. Needless to say, our formalism needs to be modified to account for other physical aspects such as spontaneous curvature, surface chemistry, or active changes of them. However, the mechanism described here, i.e., the friction-controlled buckling and slippage, is generic, and will provide a robust physical basis for understanding a range of complex biophysical problems. It may also offer an insight into the rational design of man-made products, such as those listed in the introduction, where contact friction is often unavoidable. 

\begin{acknowledgments}
We acknowledge financial support from Grants-in-Aid for Japan Society for the Promotion of Science (JSPS) Fellows (Grant~Nos.~28$\cdot$5315, to TGS) and JSPS KAKENHI (Nos.~15H03712 and~16H00815: ``Synergy of Fluctuation and Structure: Quest for Universal Laws in Non-Equilibrium Systems,'' to HW).
\end{acknowledgments}

\newpage
\section*{\normalsize Supplementary Materials}

\section{Discrete model of elastic strips}
Let us explain the discrete model of elastic strips in the main text.
The centerline of the strip is discretized into a chain of $N$ spheres with a bond length $b_0$ and a radius $\sigma \equiv b_0/2$. 
As we are interested only in the equilibrium shape, the position of each sphere, $\vec{r}_i(t) = (x_i(t),y_i(t))$ for $i=1,2,\cdots,N$, evolves according to the overdamped equations of motion with an effective friction constant $\gamma$ given by
\begin{eqnarray}
\gamma\frac{d\vec{r}_i}{dt}=\vec{F}_{i}^{\rm el}+\vec{F}_{i}^{\rm ex}+\vec{F}_{i}^{\rm s}+\vec{F}_i ^g,
\end{eqnarray}
where $\vec{F}_{i}^{\rm el}$, $\vec{F}_{i}^{\rm ex}$, $\vec{F}_{i}^{\rm s}$, and $\vec{F}_{i}^g$ are the sum of the stretching and bending elastic forces, the excluded volume force, the force from the substrate, and the gravitational force, respectively.
The elastic force $\vec{F}_{i} ^{\rm el} = \vec{F}_i ^{\rm st} + \vec{F}_i ^{\rm b}$ consists of the stretch $ \vec{F}_i ^{\rm st}$ and bending $\vec{F}_i ^{\rm b}$ forces. Internal forces acting on the $i$-th sphere are expressed as
\begin{eqnarray}
\vec{F}_i ^{\rm st} &=& -\frac{\partial U_{\rm st}}{\partial \vec{r}_i}\nonumber\\
&=& -k_{\rm st} \left[(b_{i-1}-b_0)\hat{u}_{i-1} - (b_i - b_0)\hat{u}_i\right],\\
\vec{F}_i ^{\rm b} &=& -\frac{\partial U_{\rm b}}{\partial \vec{r}_i}\nonumber\\
&=& -k_{\rm b} \left[\vec{A}_i - \vec{A}_{i-1} + \vec{B}_{i}-\vec{B}_{i-1}\right],\\
\vec{F}_i ^{\rm ex} &=& -\sum_{|j-i|\geq2} k_{\rm ex} (|\vec{r}_i - \vec{r}_j| - 2\sigma) \Theta(2\sigma - |\vec{r}_i - \vec{r}_j| )\hat{n}_{ij}\nonumber,\\
\end{eqnarray}
where we have introduced the corresponding potentials $U_{\rm st}$ and $U_{\rm b}$ and vectors $\vec{A}_i$ and $\vec{B}_i$ as 
\begin{eqnarray}
U_{\rm st} &\equiv& \frac{k_{\rm st}}{2}\sum_{i} (b_i - b_0)^2,\\
U_{\rm b} &\equiv& \frac{k_{\rm b}}{2}\sum_{i} \beta_i ^2,\label{bend_energy_model}\\
\vec{A}_i &\equiv& \frac{\beta_i}{b_i \sin{\beta_i}}(\hat{u}_{i+1}-\cos\beta_i\hat{u}_i),\\
\vec{B}_i &\equiv& \frac{\beta_{i-1}}{b_i \sin{\beta_{i-1}}}(\hat{u}_{i-1}-\cos\beta_{i-1}\hat{u}_i).
\end{eqnarray}
Here, we have introduced the distance between adjacent sphere (bond length) $b_i \equiv |\vec{r}_{i+1} - \vec{r}_i|$, the angle of adjacent bonds $\cos\beta_i \equiv \hat{u}_{i+1}\cdot\hat{u}_i$, the unit vector $\hat{u}_i \equiv (\vec{r}_{i+1}-\vec{r}_i)/b_i$, and the unit normal vector between $i$-th and $j$-th sphere $\hat{n}_{ij} \equiv (\vec{r}_i - \vec{r}_j)/|\vec{r}_i - \vec{r}_j|$. Note that the vectors $\vec{A}_i$ and $\vec{B}_i$ can be defined only if any adjacent spheres exist. $\Theta(x)$ represents the Heaviside function with $\Theta(x) = 1$ for $x > 0$ and $\Theta(x) = 0$ for $x \leq 0$.

The force from the substrate $\vec{F}^{\rm s}_i = (T_i,N_i)$ is calculated as follows. We calculate the normal force from the substrate $N_i = N_i(y_i)$, using the Lenard-Jones (LJ) potential $U_{\rm LJ}(y) \equiv \epsilon_{\rm LJ}\left({\sigma^{12}}/{y^{12}} - {\sigma^6}/{y^6}\right)$, as
\begin{eqnarray}
N_i &\equiv& -\Theta(y_c-y_i)\frac{\partial U_{\rm LJ}(y_i)}{\partial {y}_i}\nonumber\\
&=& \Theta(y_c-y_i)\frac{\epsilon_{\rm LJ}\sigma^{12}}{y_i ^{13}}\left(12-6\frac{y_i ^6}{\sigma^6}\right),
\end{eqnarray}
where $y_c\equiv 2^{1/6}\sigma \simeq 1.12\sigma$ represents the cut-off. 
Let the $j$-th sphere be in contact with the substrate ($y_j < y_c$).  
Knowing the horizontal component of the internal force acting on the $j$-th sphere, $\tilde{T}_j  \equiv\hat{e}_x\cdot(\vec{F}_{j}^{\rm el}+\vec{F}_{j} ^{\rm ex})$ with the unit vector in $x$-direction $\hat{e}_x$, $T_j$ is determined as
\begin{eqnarray}
T_j=\left\{ 
\begin{array}{ll}
-\tilde{T}_{j}& \mbox{if}\quad|\tilde{T}_{j}|\leq{\mu}N_j\\
-\mu_{\rm k}N_{j}{\rm sgn}(\tilde{T}_{j})&\mbox{if}\quad|\tilde{T}_{j}|>\mu N_j,\\
\end{array}\right.
\label{eq:CA-discrete}
\end{eqnarray}
where $\mu$ and $\mu_{\rm k}$ are the static and kinetic friction coefficients, respectively, and ${\rm{sgn}}(x)$ represents the sign of $x$.
If the $i$-th sphere is above the substrate $y_i \geq y_c$, we set ${\vec{F}^{\rm s}_{i}=\vec{0}}$.
The gravitational force is introduced as $\vec{F}_i ^g = - mg\vec{e}_y$. Here, the mass of the particle is $m$ and that of the strip is $M = mN$. The mass density of the strip per a unit length is written as $\rho = M/L$.

We prepare the straight strip $0\leq x_i(0)\leq\delta_x$ and $y_i(0) = \sigma + (i-1) b_0$ for $i = 1,2,\cdots, N$, as the initial condition. To induce initial buckling, we initially give very small noise for $x_i$, where the maximum value is set to be sufficiently small as $\delta_x/b_0 \equiv 1.0\times10^{-4}\ll 1$, except the following three spheres $i = 1, N-1,N$. We adopt $x_1(0) = 0$ and choose $x_{N-1}(0) = x_N(0) = 0$ so that the top of the strip is clamped. The clamped-end spheres $i = N-1,N$ are moved downwards at speed $u$ as $dy_{N-1}/dt = dy_N/dt = - u < 0$. We choose the units of length, velocity, and force as $2b_0$, $u$, and ${\gamma}u$, respectively, and rescale all the quantities and the equations appropriately. We integrated the rescaled equations using the Adams--Bashforth method with a rescaled time step $\delta_t = 10^{-7}$ to ensure adequate numerical accuracy. The total number of simulation time steps is $10^6-10^8$, which is sufficient for the system to reach mechanical equilibrium.

Let us explain the parameters adopted in the simulation. The strip is pushed onto the substrate sufficiently slowly so that the typical bending force is much larger than the pushing force satisfying ${k_{\rm b} b_0}/{L^2\gamma u} \gg 1$, where we adopt $k_{\rm b} = 9.0\times 10^4\gamma u b_0$ in the simulation. 
We choose $k_{\rm st}$ so that we can adopt the Kirchhoff equations, where the ratio of stretch energy and bending energy should be sufficiently large:
$k_{\rm st} L ^2/k_{\rm b} \gg 1.$
We fix $k_{\rm st} = 2.25\times 10^5\gamma u/b_0$, $k_{\rm ex} = 5.0\times10^2 \gamma u/b_0$, and $\epsilon_{\rm LJ} = \gamma ub_0/12$, where $N_i(y_i = \sigma) = 6\epsilon_{\rm LJ} /\sigma = \gamma u$ holds. 

\section{Validity of discrete strip models}

\begin{figure}[t]
\begin{center}
\includegraphics[scale = 0.70]{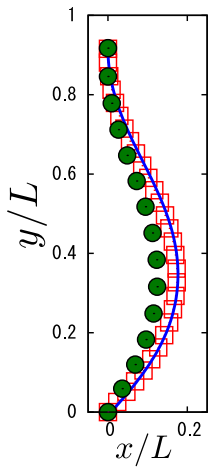}
\caption{(Color online).~Bond length dependence of discrete models.
Open squares and filled circles respectively represent the P shape of $N = 30$ and $N = 15$ with fixed $L$. The solid line is the shape constructed from continuum model.}
\label{fig:comp}
\end{center}
\end{figure}

Discrete models illustrated in the previous section is valid for small $b_0$ (large $N$) with fixed $L$, as is shown here. The pinned strip of length $L$ consisting of $N = 30$ particles and that of $N = 15$ particles are compared in Fig.~\ref{fig:comp}. We plot the shape derived from continuum model as the solid line. Apparently, the theoretical curve agrees with the results for $N = 30$ not those for $N=15$, which implies that the number of particles adopted in the main text $N = 30$ is sufficient to simulate the shape and mechanics of real elastic strips.

\section{On the modeling of kinetic friction}
\begin{figure}[b]
\begin{center}
\includegraphics[scale = 0.675]{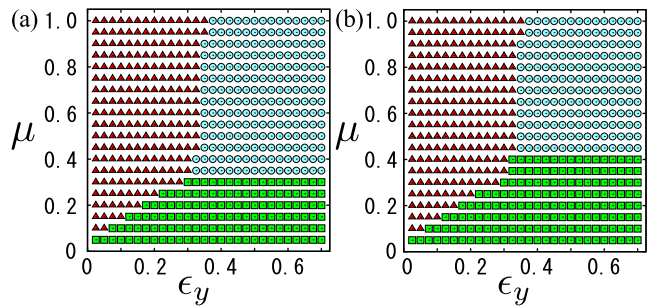}
\caption{(Color online) Phase diagrams in pushing processes for (a)~$v_{\rm th} = \delta_x/\delta_t$ and (b)~$v_{\rm th} = 0.01\delta_x/\delta_t$. Triangles, squares, circles represent P, CS, and PS states, respectively. Although the phase boundary between CS and PS states depends on $v_{\rm th}$ and CS state exists even for larger $\mu$ in (b), the phase boundaries for P are independent of the choice of $v_{\rm th}$.}
\label{fig:kin_fri}
\end{center}
\end{figure}

In some papers (e.g.~Ref.~[34] in the main text), the velocity threshold $v_{\rm th}$ is introduced in Eq.~(\ref{eq:CA-discrete}), where the static friction for $j$-th sphere sets in only if the horizontal speed of the sphere is less than the threshold speed $v_{\rm th}$ as $|dx_j/dt|<v_{\rm th}$. Using the same model of kinetic friction, we plot the phase diagram for $v_{\rm th} = \delta_x / \delta_t$ and $v_{\rm th} = 0.01\delta_x / \delta_t$ in Fig.~\ref{fig:kin_fri} (a) and (b), respectively. We find that the phase diagram for $v_{\rm th} = \delta_x/\delta_t$ is the same as the one in the main text~[Fig.~\ref{fig:kin_fri} (a)]. 
As shown in Fig.~\ref{fig:kin_fri}(b), the boundary between PS and CS states changes only slightly even for the change of $v_{\rm th}$ by the factor $10^2$, and the P-CS phase boundary remains unchanged. 
Therefore, the phase diagram is quite insensitive to the choice of the modeling of kinetic friction. 
It should be noted that the stability of P states is irrelevant of $v_{\rm th}$ because the condition of the slip is determined in a totally static manner.

\begin{figure*}[!t]
\begin{center}
\includegraphics[scale = 2.0]{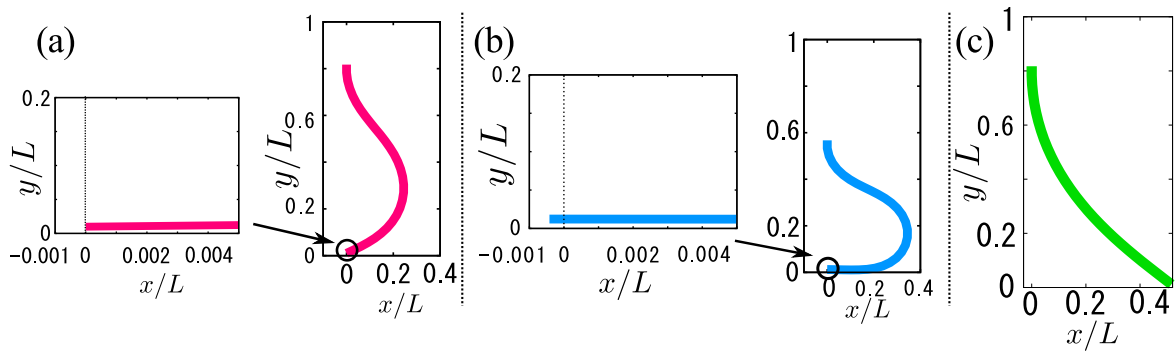}
\caption{(Color online). 
Typical configurations of (a) P state for $(\epsilon_y, \mu) = (0.2, 0.45)$, (b) PS state for $(\epsilon_y, \mu) = (0.45, 0.45)$, and (c) CS state for $(\epsilon_y, \mu) = (0.2, 0.15)$, obtained from the numerical simulations.
Close-up views of the simulated shapes around $x=0$ downside in (a) and (b) show the distinction between the P and PS states. }
\label{fig:snap_sim}
\end{center}
\end{figure*}

\section{Detailed procedure to generate the phase diagram}

We generate the phase diagram by controlling the vertical strain $\epsilon_y$ quasi-statically for fixed $\mu$. Here, we explain the protocol to generate the phase diagram. We push the strip from the height $L$ to $L(1 - \delta \epsilon_y)$ with the increment of strain $\delta \epsilon_y \equiv 0.025$ for the interval $t_{\rm push} \equiv L\delta \epsilon_y/u$. After we finish pushing for $\delta \epsilon_y$, we wait for the relaxation of the motion of the strip for the interval $t_{\rm wait} \equiv 2t_{\rm push}$ and examine the shape of the strip. Then, we push the strip from the height $L(1-\delta\epsilon_y)$ to $L(1-2\delta\epsilon_y)$ and wait for the relaxation. We repeat this procedure to reach $\epsilon_y = 0.7$ for each $\mu$.

\section{Systematic procedure to distinguish slip states}
Let us explain the systematic procedure to distinguish the shape of the strip. After the relaxation of the motion of the strip, we examine the position of the free tip, i.e., whether $|x_1|$ is smaller than $\delta_x$ or not. If $|x_1| \leq \delta_x$ holds, the strip is in the P state, otherwise, the strip state is in the PS or CS. Then, we ask the existence of an inflection point, i.e., whether $i (i = 2,3,\cdots,N-2)$ exists satisfying the condition $|x_{i+1}-x_1|-|x_{i}-x_1| < 0$ or not. If such particles exist, the strip is in the PS state. If not, the state is in the CS state. Through this procedure we can systematically generate the phase diagrams of the shape.
In Figs.~\ref{fig:snap_sim} (a), (b), and (c), we show the numerically obtained shapes for P, PS and CS states, respectively, together with the close-up views of P and PS states for distinction.

\section{Supplementary information of experiments}

\begin{figure}[h]
\begin{center}
\includegraphics[scale = 0.22]{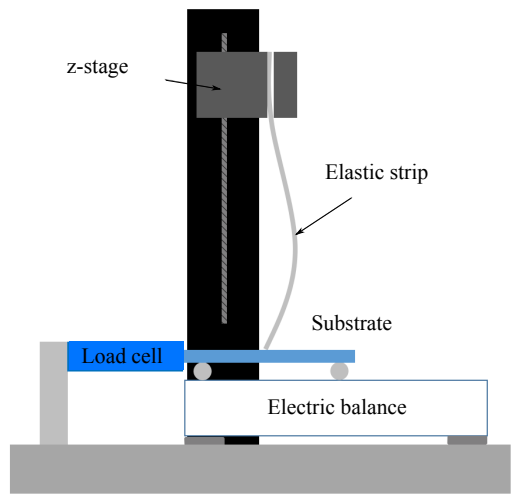}
\caption{(Color online). Schematic of the experimental setup. An elastic strip is compressed vertically 
against a substrate, and total tangential and normal forces in equilibrium are recorded by the load cell and the electric balance, respectively.}
\label{fig:sche_experiment}
\end{center}
\end{figure}

Our experimental setup is schematically depicted in Fig.~\ref{fig:sche_experiment}. An elastic strip 
made of polyvinyl chloride (PVC, $L = 150$ [mm], width: $w = 10$ [mm], thickness: $h = 1$ [mm]) 
is clamped by two metal plates which are fixed on the motorized 
$z$--stage (SGSP20-85, Sigma Koki), in which the moment of inertia $I$ is written as $I = h^3w/12 \simeq 8.33\times10^{-13}$[m$^4$]. The Young modulus $E$ of PVS ranges from 2.4 $\times10^9$ to 4.1$\times10^9$ [Pa], and the density ranges from 1.2 to 1.4 [g/cm$^3$]. Therefore, the density per unit length $\rho$ ranges from 0.012 to 0.014 [kg/m]. 
With these parameter values, the gravito-bending length $L_g = (EI/\rho g)^{1/3}$ in our experiments can be estimated from $240$ to $310$[mm], which exceeds the total length of the strip used, justifying the our assumption that the gravity is negligible compared to the bending elasticity of the strip.
To avoid plastic deformation for the strip due to the bending tendency, we change the strip of the same geometry in every pushing experiment.

The head of the $z$--stage is moved downwards at 1 mm/s by the distance of $1$-$2$\% of the strip's length. At every step, the clamping end is kept fixed about 30 seconds to attain its equilibrium configuration, and then the total tangential force ($f_x$) and normal force ($f_y$) are recorded by the load cell and the electric balance, respectively. 
The side-views are taken with a digital camera (D70S, Nikon). 
 The bottom and side faces of the PVC strip are polished with a sandpaper (\#180) 
to add some surface roughness. 
 As counterpart materials, two types of substrates 
are used, i.e., an aluminum plate (A5052P, Misumi) and a carbon-filled natural rubber sheet (thickness 1[mm]) adhered on a glass plate.

\begin{figure}[h]
\begin{center}
\includegraphics[scale = 0.8]{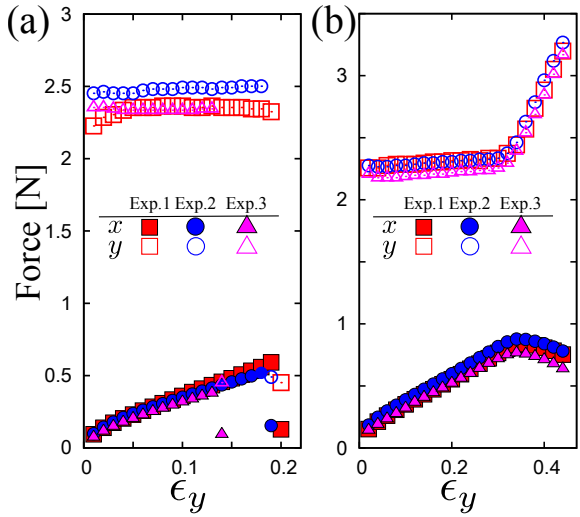}
\caption{(Color online). Real data of force-strain curves obtained from experiments with (a) aluminum and (b) rubber substrates. (a) Force-strain relations for three experiments with aluminum plates. We stop recording forces after the slip. (b) Force-strain relations from P to PS states obtained in three experiments with rubber substrates.}
\label{fig:exp_real_data}
\end{center}
\end{figure}

We have conducted three independent experiments for each substrate at room temperature. We plot those force-strain data in Fig.~\ref{fig:exp_real_data} in the physical units. 
In the experiments with the aluminum plate where the transition from P to CS was observed, the force measurement was
reliably done only before the slip, i.e. large-scale configurational change of the strip. 
At the onset of the slip, the magnitude of the force significantly drops as shown in Fig.~\ref{fig:exp_real_data}(a). 
In the main text, we show the data of experiment no.2 (Exp.2, empty and filled circles) and scale them by $EI/L^2 \simeq0.13$~[N] with the Young modulus $E = 3.5\times10^9$ [Pa]. 
In the experiment with the rubber substrate, we have performed the measurement up to $\epsilon_y = 0.44$ and, in the main text, we show the data of experiment no.3 (Exp.3, empty and filled triangles) in Fig.~\ref{fig:exp_real_data}(b), where we adopt $E = 2.7\times10^9$ [Pa] to scale the experimental data by $EI/L^2\simeq0.10$~[N].
The photographs shown in the main text correspond to those of Exp.2 and Exp.1 for aluminum and rubber substrates, respectively. 
To plot the data points in the phase diagram~[Fig.2 in the main text], we have used the data of Exp.2 and 3 for aluminum substrates.

\section{Detailed derivation of Eq.~(3) in the main text} 
In this section, we show the derivation of Eq.~(3) in the main text. Introducing $\tau\equiv s/L$, $\vartheta(\tau) \equiv \theta(L\tau)$, and $f_{\alpha} = {\mathcal F}_{\alpha}L^2/EI$ with $\alpha = x, y$, we rewrite the Kirchhoff equations in the main text as
\begin{eqnarray}
\vartheta''(\tau) &=& -f_x \cos\vartheta(\tau) - f_y\sin\vartheta(\tau)\nonumber\\
&\simeq& -f_x -f_y\vartheta\label{vartheta_eq}.
\end{eqnarray}
for $|\vartheta|\ll 1$. The solution of Eq.~(\ref{vartheta_eq}) with the boundary conditions $\vartheta(0) = 0$ and $\vartheta'(1) = 0$ is written as
\begin{eqnarray}
\vartheta(\tau) = \frac{f_x}{f_y} \left\{\tan\left(\!\sqrt{f_y}\right) \sin\left(\!\sqrt{f_y}\tau\!\right)\! + \cos\left(\!\sqrt{f_y}\tau\!\right)\!-\!1\right\}\label{sol0}.\nonumber\\
\end{eqnarray}
Because Eq.~(\ref{vartheta_eq}) is the linearized equation, $f_x/f_y$ in Eq.~(\ref{sol0}) is not determined without introducing nonlinearity of $\vartheta$. On the basis of the constraints $x(L) = 0$ and $y(L) = 0$, we can take into account the nonlinearity as
\begin{eqnarray}
0 &=& \frac{x(L)}{L} = \int_0 ^1 \sin\vartheta(\tau) d\tau \simeq  \int_0 ^1 \vartheta(\tau) d\tau,\label{x_constraint_0}\\
0 &=& \frac{y(L)}{L} = \frac{y_0}{L} - \int_0 ^1 \cos\vartheta(\tau) d\tau\nonumber\\
&\simeq& -\epsilon_y +  \frac{1}{2}\int_0 ^1 {\vartheta^2(\tau)} d\tau\label{y_constraint_0}.
\end{eqnarray}
From Eq.~(\ref{x_constraint_0}), we obtain the equation for $f_y$ as $\tan\sqrt{f_y} = \sqrt{f_y}$. And, from Eq.~(\ref{y_constraint_0}) and $f_x/f_y = \mu_c$, we also obtain
\begin{eqnarray}
\epsilon_y &\simeq& \frac{1}{2}\left(\frac{f_x}{f_y}\right)^2\left(\frac{\tan^2\sqrt{f_y}}{2} + \frac{3}{2}\left(1 - \frac{\tan\sqrt{f_y}}{\sqrt{f_y}}\right)\right)\nonumber\\
&=& \frac{f_y}{4}\mu_c ^2,
\end{eqnarray}
which corresponds to Eq.~(3) in the main text.

\begin{figure}[b]
\begin{center}
\includegraphics[scale = 0.85]{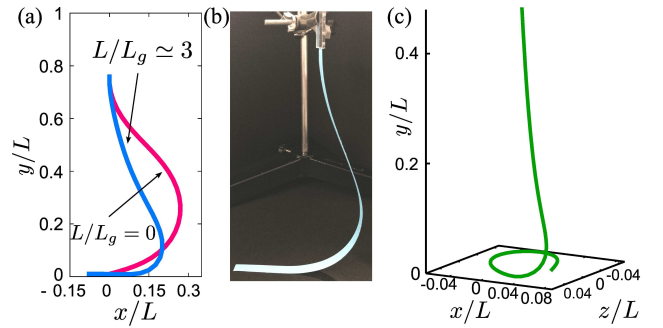}
\caption{(Color online). (a) Comparison of gravity-dominated strip ($L/L_g\simeq3$) and elasticity-dominated one ($L/L_g=0$) obtained from our simulations. Other parameters are chosen to be the same. (b) Realization of a sagging strip in our table-top experiment with a soft paper strip for illustration. (c) coiling shape in three-dimensional simulations of a twistless elastic string, for $L/L_g\simeq6$ and $N=100$.}
\label{gravity_fig}
\end{center}
\end{figure}

\section{Protocol for cycle processes}

In this section, we explain the protocol for pulling processes (Fig.~4(d) in the main text). We push the strip for the interval $t_{\rm push} = L\epsilon_y/u$ and wait for the relaxation in the same interval $t_{\rm wait} = t_{\rm push}$ with $(\epsilon_y, \mu) = (0.25, 0.20)$. After the relaxation, we bring back the strip to the straight configuration for the interval $t_{\rm push}$ by inverting the velocity of the clamped end as $u \to -u$. After pulling back the clamped end to the initial position, we wait for the relaxation for the same interval $t_{\rm wait}$.

The protocol to generate the phase diagram for the pulling processes (Fig.~4(e) in the main text) is the followings. For fixed $\mu$, we push the strip up to $\epsilon_y = 0.7$. Then, we pull the strip back to the straight configuration quasi-statically. The strain is changed from $\epsilon_y$ to $\epsilon_y - \delta\epsilon_y$ with $\delta\epsilon_y = 0.025$. After the relaxation of the strip motion, we repeat the change of the strain until the strip becomes straight: $\epsilon_y = 0.0$.

\section{Supplementary information of planer coiling states}

Let us show simulation results associated with planer coiling states. 
We study the strip shapes by changing $L/L_g$ while fixing $(\epsilon_y, \mu)=(0.15, 0.2)$ so that the strip is in the P configuration for $g=0$. 
As $L/L_g$ is increased, the strip tends to sag. The numerically obtained shape for $L/L_g \simeq 3$ and its realization of sagging strips in table-top experiments for illustration are shown in Figs.~\ref{gravity_fig}(a) and (b), respectively. As $L/L_g$ becomes larger, as shown in the main text, the strip folds and loops back on itself. Using the same parameter set where we find planer coiling states, we can reproduce a realistic coiling shape in our three-dimensional simulation of a twist-less elastic string, as shown in Fig.~\ref{gravity_fig}(c).


\begin{thebibliography}{2}
\bibitem{comment2} Inertial effects such as wave propagation, or the dynamic release of an elastic stress upon slippage will be explored elsewhere, but do not change our main results.
\bibitem{comment} However, our simulations also suggest that the force curve in the PS regime is almost independent of $\mu$, precluding any reliable estimation of the experimental value of $\mu$ in Fig.~\ref{fig:force_strain_relation} (b). A creep-like slip due to the areal contact underlies this $\mu$-indepedence, but its detailed analysis is beyond the scope here and will be published in Ref.~\cite{inclined_paper}.
\end{thebibliography}
\end{document}